\documentclass[prb,preprint]{revtex4}
\usepackage{graphicx,dcolumn,amsmath,bm}
\usepackage{pstricks}

\begin{document}

\title { 
A simple, efficient, and general treatment of the singularities in Hartree-Fock
and exact-exchange Kohn-Sham methods for solids
}

\author{\firstname{Pierre} \surname{Carrier}} 
\author{\firstname{Stefan} \surname{Rohra}}
\author{\firstname{Andreas} \surname{G\"orling}}
\affiliation{Lehrstuhl f\"ur Theoretische Chemie, Universit\"at Erlangen-N\"urnberg, Egerlandstrasse 3, 91058 Erlangen, Deutschland}

\date{\today}

\begin{abstract}
We present a general scheme for treating the integrable singular terms within exact exchange (EXX) Kohn-Sham or Hartree-Fock (HF) 
methods for periodic solids.
We show that the singularity corrections for treating these divergencies depend only on the total number and the positions of \textbf{k} points 
and on the lattice vectors, in particular
the unit cell volume, but not on the particular positions of atoms within the unit cell.
The method proposed here to treat the singularities constitutes a stable, simple to implement, and general scheme that can be applied to systems with arbitrary lattice parameters 
  within either the EXX Kohn-Sham or the HF formalism.
We apply the singularity correction to a typical symmetric structure, diamond, and to a more general structure, \textit{trans}-polyacetylene.
We consider the effect of the singularity corrections on volume optimisations and \textbf{k} point convergence.
While the singularity corrections clearly depends on the total number of \textbf{k} points, it exhibits a remarkably small dependence upon the choice of the specific arrangement of 
  the \textbf{k} points.
\end{abstract}

\pacs{71.15.Mb, 71.45.Gm, 71.55.Ak}

\maketitle
\section{Introduction}
In recent years exact exchange (EXX) Kohn-Sham (KS) methods for solids became increasingly popular\cite{Staedele, Neugebauer, Schaeffler, Magyar,StaedelePRL, Ambrosch} as alternative
  to conventional KS procedures based on the local density approximation\cite{Kohn, LDA} (LDA) or generalised gradient approximations (GGAs).\cite{GGA} 
EXX-KS methods treat both the exchange energy as well as the local KS exchange potential, not to be confused with the non-local Hartree-Fock exchange potential, exactly. 
This means they constitute a systematic improvement over LDA and GGA methods in the sense that only the correlation energy and potential, i.e., contributions of higher order in the
  electron-electron interaction, need to be approximated whereas the terms of first order in the square $e^2$ of the electron charge, i.e., the Coulomb and exchange energy and potential, are treated exactly.\cite{JCPGoerling}
Because exchange and Coulomb potential and energy are treated exactly unphysical self-interactions contained in the Coulomb energy and potential are completely cancelled by the exchange energy and potential.
EXX methods therefore are free of Coulomb self-interactions. 
As a result EXX band structures and in particular band gaps are strongly improved compared to those from LDA or GGA methods. 
Indeed for medium gap semiconductors EXX methods yield band gaps\cite{Staedele} which are very close to the experimental ones\cite{Madelung} despite the fact that the correlation potential needs to be neglected or
  approximated by conventional LDA or GGA functionals, and despite the fact that the KS band gap does not account for the derivative discontinuity\cite{PerdewDISC, Sham} of the band gap at integer electron numbers.

A second first-principles approach besides the family of density-functional methods is the Hartree-Fock (HF) method.\cite{HFcrystal, Massidda} 
Recently, there has been an increasing interest in HF methods for solids as basis for higher level approaches like, e.g., M{\o}ller-Plesset,\cite{Sun, Ayala} coupled cluster\cite{Schuetz} 
or multireference configuration interaction\cite{Birkenheuer} methods.

Both in the EXX and in the HF formalism the exchange energy contains divergent terms.\cite{Gygi} 
In the limit of an infinite system, i.e., the limit of an infinite number of $\textbf{k}$ points, the divergencies are integrable. 
In this limit the exchange energy is therefore well-defined. 
Moreover, corresponding divergencies also occur in the matrix elements of the non-local exchange operator which is required in the HF self-consistency process and can be used in the construction of the 
  local KS exchange potential.\cite{StaedelePRL, Staedele} 
Also here the divergencies are integrable in the limit of an infinite number of $\textbf{k}$ points. 
The question arises how to treat these divergencies in practical calculations which necessarily take into account only a finite number of $\textbf{k}$ points. 
Indeed, in order to keep the computational effort as low as possible, it is preferable to keep the number of $\textbf{k}$ points as low as possible. 
This, however, is possible only if an adequate treatment of the singularities is available. 
Moreover, such a treatment of the singularities should be computationally efficient and ideally its implementation should not require much programming effort. 
Gygi and Baldereschi \cite{Gygi} presented such a method for the case of zincblende (fcc) structures.
Wenzien et al.\cite{Wenzien} further generalised the method to simple cubic, bcc, hexagonal, and orthorhombic structures. 
For other crystal structures such simple and straightforward method, to our knowledge, is still lacking 
and alternative approaches \cite{Massidda,Sorouri,Furthmueller} are more 
involved. In Refs. [\onlinecite{Massidda,Sorouri}], e.g., a general treatment of the singularities is 
presented. This method, however, is somewhat laborious because it requires a quadrature over reciprocal lattice vectors at each \textbf{k} points.
This quadrature formally has to run over an infinite number of reciprocal lattice vectors which in practice needs to be approximated by a finite summation.
Thus there is demand for a simple, efficient treatment of the singularities, that is applicable to {\em arbitrary} crystal structures. 

In this article we present a \emph{simple, efficient, and general} treatment of the singularities in Hartree-Fock and exact-exchange Kohn-Sham methods for periodic systems, which extends the approach of
Gygi and Baldereschi\cite{Gygi} to systems with arbitrary lattice parameters. 
The derivation of this treatment of the singularities is accompanied by an analysis of the singularities and demonstrates the simplicity of the suggested method for handling these singularities. 
In order to demonstrate the applicability of the approach we present results for the symmetric diamond (fcc) structure (2 carbon atoms) as well as for \textit{trans}-polyacetylene 
  (4 carbons and 4 hydrogens in its crystalline unit cell).
Polyacetylene has monoclinic symmetry $P2_1/a$, i.e., non-orthogonal lattices,\cite{Vogl} and constitutes a simple example of an organic polymer.

The article is organized as follows. In Section \ref{formEXX} the general treatment of the divergent terms in EXX and HF methods is derived and discussed.
Section \ref{Results} unwraps the results for diamond and \textit{trans}-polyacetylene.
Section \ref{Conclude} concludes.

\section{Total energy within the EXX formalism}
\label{formEXX}

We start by considering the expressions for the total electronic energy $E_0$ within the Kohn-Sham (KS) and the Hartree-Fock (HF) schemes. 
Within the KS formalism the total ground state energy is decomposed into the non-interacting kinetic energy $T_s$, the Coulomb energy $U$,
the exchange energy $E_x$, the correlation energy $E_c$, and the
  interaction energy with the external potential, $V_{ext}$:\cite{Kohn}
\begin{equation}
E_0 = T_s + U + E_x + E_c + \int \! d\textbf{r} \, V_{ext}(\textbf{r})\rho_0(\textbf{r}).
\end{equation}
The non-interacting kinetic energy $T_s$ is evaluated exactly via the KS orbitals. 
The contributions $U$ and $\int d\textbf{r} V_{ext}(\textbf{r})\rho_0(\textbf{r})$ also can be calculated exactly for a
  given electron density and thus also for the ground state electron density $\rho_0$. 
The correlation energy $E_c$ in almost all KS schemes is evaluated approximately within the LDA\cite{LDA} or the GGA.\cite{GGA} 
The exchange energy $E_x$ can either be evaluated via the LDA or the GGA within a conventional KS scheme, or exactly within the EXX-KS scheme.\cite{JCPGoerling}
 
The HF total energy, on the other hand, is decomposed into 
\begin{equation}
E_0 = T + U + E_x + \int \! d\textbf{r} \, V_{ext}(\textbf{r})\rho_{HF}(\textbf{r}).
\end{equation}
Within HF schemes all contributions of the energy are usually treated exactly: the kinetic energy $T$ and the exchange energy $E_x$ via the HF orbitals, 
and $U$ and $\int d\textbf{r} V_{ext}(\textbf{r})\rho_{HF}(\textbf{r})$ 
  via the HF electron density $\rho_{HF}$. 

The exact-exchange energy $E_x$ per unit cell for a crystalline solid, either for the KS or HF formalisms, is given by:
\begin{eqnarray}
E_x = 
-\frac{1}{N_k} \sum_{v\textbf{k}}^{\mbox{occ.}}
\sum_{w\textbf{q}}^{\mbox{occ.}} \int_{\Omega} \! d\textbf{r} 
\int_{\Omega} \! d\textbf{r}^{'} \,
\frac{\phi^{\dag}_{v\textbf{k}}(\textbf{r}) \phi_{w\textbf{q}}(\textbf{r})
      \phi^{\dag}_{w\textbf{q}}(\textbf{r}^{'}) \phi_{v\textbf{k}}(\textbf{r}^{'})}
{|\textbf{r}-\textbf{r}^{'}|}, 
\label{ExSigma}
\end{eqnarray}
where both summations run through all occupied single-particle wave functions, i.e., 
orbitals $\phi_{v\textbf{k}}$ and $\phi_{w\textbf{k}}$ for each \textbf{k} point in the Brillouin zone (BZ).
All orbitals are assumed to be normalised with respect to the crystal volume $\Omega = N_k V$ where 
$V$ designates the volume of the unit cell, and $N_k$ denotes the number of \textbf{k} points.
We implicitely treat the spin via appropriate prefactors in summations and consider for simplicity non-spin polarised calculations.
The Coulomb interaction term, $\frac{1}{|\textbf{r}-\textbf{r}^{'}|}$ in Eq.\ (\ref{ExSigma}), has to be treated taking into account periodic boundary conditions.
Note that, despite the fact that the expression for the exchange energy in terms of one-particle functions is identical in the KS and HF case, 
the KS and HF exchange energies remain
  different because their respective one-particle functions are constructed using two different scheme: 
KS uses a \emph{local} exchange operator while HF uses a \emph{non-local} exchange operator.

After expressing the \emph{product} of one-particle functions as
\begin{equation}
\phi^{\dag}_{w\textbf{q}}(\textbf{r})\phi_{v\textbf{k}}(\textbf{r}) = \frac{1}{\Omega} \sum_{\textbf{G}}
Y_{w\textbf{q},v\textbf{k}}(\textbf{G}) e^{i(\textbf{G}+\textbf{k}-\textbf{q})\cdot\textbf{r}},
\label{Ysum}
\end{equation}
with 
\begin{equation}
Y_{w\textbf{q},v\textbf{k}}(\textbf{G}) = 
\int_{\Omega} \! d\textbf{r} \; e^{-i(\textbf{G}+\textbf{k}-\textbf{q})\cdot\textbf{r}} \,
\phi^{\dag}_{w\textbf{q}}(\textbf{r}) \, \phi_{v\textbf{k}}(\textbf{r})
\label{Ywv}
\end{equation}
one obtains the following expression for the exchange energy per unit cell
\begin{equation}
E_x  =
-\frac{4\pi}{N_k\Omega} \sum_{v\textbf{k}} \sum_{w\textbf{q}} \sum_{\textbf{G}}
\frac{Y^{*}_{w\textbf{q},v\textbf{k}}(\textbf{G}) \, 
Y_{w\textbf{q},v\textbf{k}}(\textbf{G})}{|\textbf{G}+\textbf{k}-\textbf{q}|^2},
\label{ex1}
\end{equation}
if the following relation is taken into account
\begin{equation}
\int_{\Omega} \! d\textbf{r} \int_{\Omega} \! d\textbf{r}^{'} \, 
\frac{e^{-i \textbf{G} \cdot \textbf{r}} \, e^{i \textbf{G}^{'} \cdot \textbf{r}^{'}}}
{|\textbf{r} - \textbf{r}^{'}|} 
 = 
\frac{4\pi \Omega}{|G|^2} \; \delta_{\textbf{G}\textbf{G}^{'}},
\label{Gsquare}
\end{equation}
which holds due to translational symmetry. 

Expression (\ref{ex1}) contains singular terms, namely those with $\textbf{G}=\textbf{0}$, $\textbf{k} = \textbf{q}$, and $v=w$. 
Note that when $v\ne w$ no singularities occur for \emph{any} value of $\textbf{G}$ and $\textbf{k}$.
This is due to the relation
\begin{equation}
Y_{w\textbf{k},v\textbf{k}}(\textbf{0}) = \delta_{wv}
\label{nullY}
\end{equation}
which holds because Eq.\ (\ref{Ywv}) that defines $Y_{w\textbf{q},v\textbf{k}}(\textbf{0})$ in the case where $\textbf{G}=\textbf{0}$ and $\textbf{k} = \textbf{q}$ just turns into the orthonormality condition for the 
  one-particle functions. 
Thus, contributions with $\textbf{G}=\textbf{0}$, $\textbf{k} = \textbf{q}$, and $v \ne w$ vanish because the plane wave representations of the products $\phi^{\dag}_{w\textbf{k}}(\textbf{r}) \phi_{v\textbf{k}}(\textbf{r})$ 
  with $v \ne w$ do not contain contributions from a plane wave with $\textbf{G}=\textbf{0}$. 
[This means that for $v\ne w$ no singularities are present in Eq.\ (\ref{ex1}). 
Therefore, stricktly speaking, Eq.\ (\ref{ex1}) needs to be modified in a way that for $v \ne w$ singular terms are no longer present]. 

Due to the presence of the singularities described above the exchange energy is well-defined only in the limit of an infinite number of unit cells, i.e., for $N_k \rightarrow \infty$. 
In this case, the summations over $\textbf{k}$ and $\textbf{q}$ turn into integrals over the BZ and Eq.\ (\ref{ex1}) for the exchange energy assumes the form
\begin{eqnarray}
E_x  =
-\frac{4\pi}{N_k \Omega} \frac{\Omega^2}{(2\pi)^6} \sum_{v}  \int_{BZ} \! d\textbf{k} \, 
\sum_{w} \int_{BZ} \! d\textbf{q} \, \sum_{\textbf{G}}
\frac{Y^{*}_{w\textbf{q},v\textbf{k}}(\textbf{G}) \, 
Y_{w\textbf{q},v\textbf{k}}(\textbf{G})}{|\textbf{G}+\textbf{k}-\textbf{q}|^2}.
\label{ex2}
\end{eqnarray}
The singularities in integral (\ref{ex2}) are integrable. 
Therefore, the exchange energy is now well-defined.

Adopting an idea of Gygi and Baldereschi\cite{Gygi} we now manipulate the contributions on the right hand side of Eq.\ (\ref{ex2}) with $\textbf{G}=\textbf{0}$ and $v=w$, 
  i.e., those contributions which contain the integrable singularities, by adding and subtracting a function $f(\textbf{q})$ which shall obey the three following conditions:
(i) $f(\textbf{q})$ is periodic within the reciprocal lattice, 
(ii) $f(\textbf{q})$ diverges as $1/q^2$ for $\textbf{q} \rightarrow \textbf{0}$ and is smooth elsewhere, and 
(iii) $f(\textbf{q}) = f(-\textbf{q})$.  
This leads to
\begin{eqnarray}
&& \!\!\!\!\!\!\!\!-\frac{4\pi}{N_k\Omega} \frac{\Omega^2}{(2\pi)^6} \sum_{v}  \int_{BZ} \! d\textbf{k} \,
\int_{BZ} \! d\textbf{q} \,
\frac{Y^{*}_{v\textbf{q},v\textbf{k}}(\textbf{0}) \, 
Y_{v\textbf{q},v\textbf{k}}(\textbf{0})}{|\textbf{k}-\textbf{q}|^2}
\nonumber \\
&=& -\frac{4\pi}{N_k\Omega} \frac{\Omega^2}{(2\pi)^6} \sum_{v} \int_{BZ} \! d\textbf{k} \,
 \int_{BZ} \! d\textbf{q} \, \bigg[
\frac{Y^{*}_{v\textbf{q},v\textbf{k}}(\textbf{0}) \, 
Y_{v\textbf{q},v\textbf{k}}(\textbf{0})}{|\textbf{k}-\textbf{q}|^2} 
-f(\textbf{k}-\textbf{q}) \bigg] 
\nonumber \\
&& -\frac{4\pi}{N_k\Omega} \frac{\Omega^2}{(2\pi)^6} \sum_{v} \int_{BZ} \! d\textbf{k} \,
 \int_{BZ} \! d\textbf{q} \; f(\textbf{k}-\textbf{q})
\nonumber \\
&\approx& -\frac{4\pi}{N_k\Omega} \sum_{v} \sum_{\textbf{k}}
\sum_{\textbf{q} \ne \textbf{k}} \bigg[
\frac{Y^{*}_{v\textbf{q},v\textbf{k}}(\textbf{0}) \, 
Y_{v\textbf{q},v\textbf{k}}(\textbf{0})}{|\textbf{k}-\textbf{q}|^2} -
f(\textbf{k}-\textbf{q}) \bigg]
\nonumber \\
&& -\frac{4\pi}{N_k\Omega} \frac{\Omega^2}{(2\pi)^6} \sum_{v} \int_{BZ} \! d\textbf{k} \,
 \int_{BZ} \! d\textbf{q} \; f(\textbf{q})
\nonumber \\
&=& -\frac{4\pi}{N_k\Omega} \sum_{v} \sum_{\textbf{k}}
\sum_{\textbf{q} \ne \textbf{k}} 
\frac{Y^{*}_{v\textbf{q},v\textbf{k}}(\textbf{0}) \, 
Y_{v\textbf{q},v\textbf{k}}(\textbf{0})}{|\textbf{k}-\textbf{q}|^2}
\nonumber \\
&& +\frac{4\pi N_v}{N_k\Omega} \sum_{\textbf{k}}
\sum_{\textbf{q} \ne \textbf{k}} f(\textbf{k}-\textbf{q}) 
\;-\;  N_v \frac{4\pi}{(2\pi)^3} \,
 \int_{BZ} \! d\textbf{q} \; f(\textbf{q})
\nonumber \\
&=& -\frac{4\pi}{N_k\Omega} \sum_{v} \sum_{\textbf{k}}
\sum_{\textbf{q} \ne \textbf{k}} 
\frac{Y^{*}_{v\textbf{q},v\textbf{k}}(\textbf{0}) \, 
Y_{v\textbf{q},v\textbf{k}}(\textbf{0})}{|\textbf{k}-\textbf{q}|^2}
+ N_v [\tilde{F} - F],
\label{singularterms2}
\end{eqnarray}
where
\begin{equation}
\tilde{F} = \frac{1}{N_k} \sum_{\textbf{k}}
\tilde{F}_{\textbf{k}}
= \frac{1}{N_k} \sum_{\textbf{k}}
\left[\frac{4\pi }{\Omega} \sum_{\textbf{q} \ne \textbf{k}} f(\textbf{k}-\textbf{q})\right]
\label{Fsum}
\end{equation}
and 
\begin{equation}
F =\frac{4\pi}{(2\pi)^3} \, \int_{BZ} \! d\textbf{q} \; f(\textbf{q}).
\label{Finteg}
\end{equation}
The function $\tilde{F}_{\textbf{k}}$ in Eq.\ (\ref{Fsum}) is given by
\begin{equation}
\tilde{F}_{\textbf{k}}
=\frac{4\pi }{\Omega} \sum_{\textbf{q} \ne \textbf{k}}
f(\textbf{k}-\textbf{q}) \, .
\label{Fsumk}
\end{equation}

In Eq.\ (\ref{singularterms2}) $N_v$ designates the number of valence bands, which comes from the summation over the valence bands in the two terms containing the function $f$.
We also used condition (ii) for the function $f$ and Eq.\ (\ref{nullY}) that require
  ${Y^{*}_{v\textbf{q},v\textbf{k}}(\textbf{0}) \, Y_{v\textbf{q},v\textbf{k}}(\textbf{0})}/{|\textbf{k}-\textbf{q}|^2} - f(\textbf{k}-\textbf{q})$ for any given $\textbf{k}$ 
  to be a smooth function of $\textbf{q}$ that equals zero at $\textbf{q} = \textbf{k}$.
Therefore, the first integral over $\textbf{q}$ and $\textbf{k}$ after the first equality sign of Eq.\ (\ref{singularterms2}) can be evaluated by summations over the finite grid of $\textbf{k}$ points omitting the 
  terms with $\textbf{q} = \textbf{k}$.
Due to the periodicity and inversion symmetry of $f(\textbf{q})$ [condition (i) and (iii) for $f$] the integrals of $f$(\textbf{k}-\textbf{q}) over the BZ can be replaced by BZ integrals of $f(\textbf{q})$. 
Furthermore, for the case of a uniform grid of $\textbf{k}$ points the function $\tilde{F}$ of Eq.\ (\ref{Fsum}) simplifies to 
\begin{equation}
\tilde{F} = \frac{4\pi}{\Omega} \sum_{\textbf{q} \ne \textbf{0}} f(\textbf{q}).
\label{uniform}
\end{equation}

The evaluation of the exchange energy can now be done according to 
\begin{eqnarray}
E_x  &=&
-\frac{4\pi}{N_k\Omega} \sum_{v,\textbf{k}} \; \sum_{w,\textbf{q}\ne\textbf{k}} \sum_{\textbf{G}}
\frac{Y^{*}_{w\textbf{q},v\textbf{k}}(\textbf{G}) \, 
Y_{w\textbf{q},v\textbf{k}}(\textbf{G})}{|\textbf{G}+\textbf{k}-\textbf{q}|^2}
\nonumber \\
&&
-\frac{4\pi}{N_k\Omega} \sum_{v,\textbf{k}} \sum_{w} \sum_{\textbf{G}\ne \textbf{0}}
\frac{Y^{*}_{w\textbf{k},v\textbf{k}}(\textbf{G}) \, 
Y_{w\textbf{k},v\textbf{k}}(\textbf{G})}{|\textbf{G}|^2}
\nonumber \\
&&+ N_v [\tilde{F} - F] \,.
\label{ex3}
\end{eqnarray}
This implies that for the evaluation of the exchange energy the singular terms in the original expression (\ref{ex1}) can first simply be omitted and then be taken into account by $N_v [\tilde{F} - F]$,
  i.e., by adding $N_v [\tilde{F} - F]$ to the exchange energy obtained if the singular terms are simply omitted.
The correction is calculated only once before the self-consistency procedure. 
In fact, the correction $N_v [\tilde{F} - F]$ depends only on the unit cell lattice vectors, and thus in particular on the unit cell volume $V$, and on the number $N_k$ and the positions of the \textbf{k} points. 
It does not depend on the number, type, or positions of the atoms within the unit cell,
  and does not depend on the one-particle wave functions.
This has obvious advantages for atomic relaxations at fixed unit cell volumes and fixed lattices.

The whole scheme, of course, hinges on the availability of a suitable function $f(\textbf{q})$. 
For fcc systems such a function was given by Gygi and Baldereschi.\cite{Gygi} For sc, bcc, hexagonal, and orthorhombic systems Wenzien\cite{Wenzien} presented such functions. 
Here we suggest the following function $f$ for {\em arbitrary} crystal structures:
\begin{eqnarray}
f(\textbf{q}) & = & \frac{1}{1/(2\pi)^{2}}\left\{4\sum_{j=1}^{3} [\textbf{b}_j
       \sin(\textbf{a}_j \cdot \textbf{q}/2)] \cdot [\textbf{b}_j
      \sin(\textbf{a}_j \cdot \textbf{q}/2)]\right. 
\nonumber \\
       & + & \left. 2\sum_{j=1}^{3} [\textbf{b}_j \sin(\textbf{a}_j \cdot \textbf{q})] 
\cdot [\textbf{b}_{j+1} \sin(\textbf{a}_{j+1} \cdot \textbf{q})] \right\}^{-1}. 
 \label{funcf}
\end{eqnarray}
The $\textbf{b}_j$ (with $\textbf{b}_4\equiv\textbf{b}_1$ for a compact formulation accounting for cyclic permutations) are the reciprocal lattice vectors, and the $\textbf{a}_j$ (with
  $\textbf{a}_4\equiv\textbf{a}_1$) are the corresponding lattice vectors spanning the unit cell. 
The coefficient $1/(2\pi)^{2}$ arises from the factor $2\pi$ contained in the Taylor expansion of the trigonometric functions, because $\textbf{a}_j \cdot \textbf{q}$ implicitely contains
$\textbf{a}_j \cdot \textbf{b}_j = 2\pi$ if $\textbf{q}$ is expressed as $\textbf{q} = \sum_j q_j \textbf{b}_{j}$, with $q_j$ describing the components of $\textbf{q}$ with respect to reciprocal lattice vectors.
The function (\ref{funcf}) by construction has the required periodicity of the reciprocal unit cell. 
Expansion into a Taylor series with respect to the cartesian components $q_x$, $q_y$, and $q_z$, or equivalently with respect to $q_1$, $q_2$, and $q_3$, the components of $\textbf{q}$ referring to the reciprocal lattice,
furthermore shows that it diverges as $1/q^2$ for $\textbf{q} \rightarrow \textbf{0}$. 
Therefore, $f(\textbf{q})$ satisfies requirements (i)--(iii) for any type of (linearly independent) lattice parameters: $\textbf{a}_1$, $\textbf{a}_2$, and $\textbf{a}_3$.

The integration over the BZ of the function (\ref{funcf}) required for obtaining the correction $F$ in Eq.\ (\ref{Finteg}) can easily be carried out numerically on an adaptive grid using an iterative algorithm. 
To this end, we place the reciprocal lattice centered symmetrically around $\textbf{q}=\textbf{0}$. In the first iteration we generate a regular 
$(2N+1) \times (2N+1) \times (2N+1)$-grid
  with the number $N$ being a multiple of 3, typically $N=60$.\cite{note60} 
The grid points shall be labeled $\textbf{q}_{\ell mn}$ with $-N \le \ell \le N$, $-N \le m \le N$, and $-N \le n \le N$, with the point $\textbf{q}_{000}$ located at the origin of the reciprocal lattice. 
We then divide the unit cell into an inner part given by a cell in reciprocal space which again is centered symmetrically around $\textbf{q} = \textbf{0}$, and which is defined by lattice vectors being one third 
  of the original reciprocal unit cell vectors.
In the first iteration numerical integration is carried out only in the outer region. 
Then in the inner region, where the singularity is located, the number of points is tripled and a second iteration proceeds as the first one, working now on the outer part of the inner region of the first iteration. 
By moving on in this fashion the algorithm triples the mesh size around the singularity in each iteration step. 
Therefore, the integration result is more accurate than with any regular mesh. 
We observe that less than 10 steps, depending on the lattice vectors considered, are sufficient to get the integral converged. 
The implementation of the described numerical integration is straightforward leading to about 200 lines of \textsc{Fortran} instructions. The computational time for carrying out the integration is negligible. 

Having considered in detail the treatment of the singularities in the exchange energy we now briefly present the corresponding treatment of singularities in
  the evaluation of the matrix elements of the non-local exchange potential, which is required in the HF self-consistency process, or can be used in the
  construction of the local KS exchange potential\cite{StaedelePRL, Staedele} during the self-consistency process of a KS calculation. 

The matrix elements of the non-local exchange potential, $v_x^{NL}(\textbf{k},\mu, \nu)$, are given by
\begin{eqnarray}
v_x^{NL}(\textbf{k}, \mu, \nu) = - \sum_{w\textbf{q}} \int_{\Omega} d\textbf{r} d\textbf{r}^{'} \;
\frac{\chi_{\mu\textbf{k}}(\textbf{r}^{'}) \,  \phi_{w\textbf{q}}(\textbf{r}^{'})\phi^{\dag}_{w\textbf{q}}(\textbf{r}) \,\chi_{\nu\textbf{k}}(\textbf{r}) }{|\textbf{r}-\textbf{r}^{'}|} 
  \hspace{2.7cm} & & 
\label{VxNL}
\end{eqnarray}
with $\chi_{\mu\textbf{k}}$ and $\chi_{\nu\textbf{k}}$ denoting the basis functions
for the representation of the one-particle functions
$\phi_{w\textbf{q}}$. The basis functions
$\chi_{\mu\textbf{k}}$ are products of a periodic part and a Bloch factor
$e^{-i\textbf{k} \cdot \textbf{r}}$. The most common choice for the  basis
functions $\chi_{\mu\textbf{k}}$ are plane waves $e^{-i(\textbf{G} +
  \textbf{k}) \cdot \textbf{r}}$. Like in the treatment of the exchange energy,
we now express the products $\phi^{\dag}_{w\textbf{q}}(\textbf{r})
\,\chi_{\nu\textbf{k}}(\textbf{r})$  as 
\begin{equation}
\phi^{\dag}_{w\textbf{q}}(\textbf{r}) \,\chi_{\nu\textbf{k}}(\textbf{r}) = \frac{1}{\Omega} \sum_{\textbf{G}}
Y_{w\textbf{q},\nu\textbf{k}}(\textbf{G}) e^{i(\textbf{G}+\textbf{k}-\textbf{q})\cdot\textbf{r}},
\label{Ysum2}
\end{equation}
and obtain
\begin{equation}
v_x^{NL}(\textbf{k}, \mu, \nu) =
-\frac{4\pi}{\Omega}  \sum_{w\textbf{q}} \sum_{\textbf{G}}
\frac{Y^{*}_{w\textbf{q},\mu\textbf{k}}(\textbf{G}) \, 
Y_{w\textbf{q},\nu\textbf{k}}(\textbf{G})}{|\textbf{G}+\textbf{k}-\textbf{q}|^2},
\label{vx1}
\end{equation}
Expression (\ref{vx1}) contains singular terms, again those with
$\textbf{G}=\textbf{0}$ and $\textbf{k} = \textbf{q}$. In the limit of an
infinite number of unit cells the summation over $\textbf{q}$ again turns into
an integral, namely
\begin{eqnarray}
 v_x^{NL}(\textbf{k}, \mu, \nu)=
-\frac{4\pi}{\Omega} \frac{\Omega}{(2\pi)^3} \, 
\sum_{w} \int_{BZ} \! d\textbf{q} \, \sum_{\textbf{G}}
\frac{Y^{*}_{w\textbf{q},\mu\textbf{k}}(\textbf{G}) \, 
Y_{w\textbf{q},\nu\textbf{k}}(\textbf{G})}{|\textbf{G}+\textbf{k}-\textbf{q}|^2},
\label{vx2insulator}
\end{eqnarray}
with an integrable singularity. We can now treat the singular terms in the
right hand side of (\ref{vx2insulator}) exactly analogously to the singular terms
occuring in the exchange energy:
\begin{eqnarray}
&& \!\!\!\!\!\!\!\!-\frac{4\pi}{\Omega}  \frac{\Omega}{(2\pi)^3} \sum_{w}
\int_{BZ} \! d\textbf{q} \,
\frac{Y^{*}_{w\textbf{q},\mu\textbf{k}}(\textbf{0}) \, 
Y_{w\textbf{q},\nu\textbf{k}}(\textbf{0})}{|\textbf{k}-\textbf{q}|^2}
\nonumber \\
&=& -\frac{4\pi}{\Omega} \frac{\Omega}{(2\pi)^3} 
\sum_{w}  \int_{BZ} \! d\textbf{q} \, \bigg[
\frac{Y^{*}_{w\textbf{q},\mu\textbf{k}}(\textbf{0}) \, 
Y_{w\textbf{q},\nu\textbf{k}}(\textbf{0})}{|\textbf{k}-\textbf{q}|^2} 
\, - \,  Y^{*}_{w\textbf{k},\mu\textbf{k}}(\textbf{0}) \, 
Y_{w\textbf{k},\nu\textbf{k}}(\textbf{0})  \, f(\textbf{k}-\textbf{q}) \bigg] 
\nonumber \\
&& - \, \left [ \sum_{w}  Y^{*}_{w\textbf{k},\mu\textbf{k}}(\textbf{0}) \, 
Y_{w\textbf{k},\nu\textbf{k}}(\textbf{0}) \right ] \,\frac{4\pi}{\Omega} \frac{\Omega}{(2\pi)^3} 
 \int_{BZ} \! d\textbf{q} \; f(\textbf{k}-\textbf{q})
\nonumber \\
&\approx& -\frac{4\pi}{\Omega} \sum_{w}
\sum_{\textbf{q} \ne \textbf{k}} \bigg[
\frac{Y^{*}_{w\textbf{q},\mu\textbf{k}}(\textbf{0}) \, 
Y_{w\textbf{q},\nu\textbf{k}}(\textbf{0})}{|\textbf{k}-\textbf{q}|^2} - \, Y^{*}_{w\textbf{k},\mu\textbf{k}}(\textbf{0}) \, 
Y_{w\textbf{k},\nu\textbf{k}}(\textbf{0}) \,
f(\textbf{k}-\textbf{q}) \bigg]
\nonumber \\
&& - \, \left [ \sum_{w} Y^{*}_{w\textbf{k},\mu\textbf{k}}(\textbf{0}) \, 
Y_{w\textbf{k},\nu\textbf{k}}(\textbf{0}) \right ] \,\frac{4\pi}{\Omega} \frac{\Omega}{(2\pi)^3} 
 \int_{BZ} \! d\textbf{q} \; f(\textbf{q})
\nonumber \\
&=& -\frac{4\pi}{\Omega} \sum_{w} 
\sum_{\textbf{q} \ne \textbf{k}} 
\frac{Y^{*}_{w\textbf{q},\mu\textbf{k}}(\textbf{0}) \, 
Y_{w\textbf{q},\nu\textbf{k}}(\textbf{0})}{|\textbf{k}-\textbf{q}|^2}
\, + \, \left [ \sum_{w} Y^{*}_{w\textbf{k},\mu\textbf{k}}(\textbf{0}) \, 
Y_{w\textbf{k},\nu\textbf{k}}(\textbf{0}) \right ] \, [\tilde{F}_\textbf{k} - F],
\label{singularterms3}
\end{eqnarray}
with $\tilde{F}_{\textbf{k}} = \frac{4\pi}{\Omega} \sum_{\textbf{q}\ne \textbf{k}} f({\textbf{k} - \textbf{q}})$ and $F = \frac{4\pi}{(2\pi)^3} \int_{BZ}\! d\textbf{q}\;  f(\textbf{k} - \textbf{q})$.

Thus, the matrix elements $v_x^{Nl}(\textbf{k},\mu,\nu)$ of the non-local exchange potential
can be evaluated by first omitting the singular
terms in 
Eq.\ (\ref{vx1}) and by then adding the
following correction term
\begin{equation}
\bigg[\sum_{w} Y^{*}_{w\textbf{k},\mu\textbf{k}}(\textbf{0}) \, 
Y_{w\textbf{k},\nu\textbf{k}}(\textbf{0}) \bigg] \bigg( \tilde{F}_{\textbf{k}} - F\bigg).
\label{corrHFinsulator}
\end{equation}
The required sums $\tilde{F}_{\textbf{k}}$ and the integral $F$ have to be calculated only \emph{once} at the beginning of the self-consistency procedure and then are multiplied by 
$\sum_{w} Y^{*}_{w\textbf{k},\mu\textbf{k}}(\textbf{0}) \, 
Y_{w\textbf{k},\nu\textbf{k}}(\textbf{0})$ in each HF self-consistency cycle, because the $Y^{*}_{w\textbf{k},\mu\textbf{k}}(\textbf{0}) \, 
Y_{w\textbf{k},\nu\textbf{k}}(\textbf{0})$ change
\emph{during} the self-consistency cycle. 
In case of a uniform grid of $\textbf{k}$ points the $\tilde{F}_{\textbf{k}}$ reduce to $\tilde{F}$ given in Eq.\ (\ref{uniform}).

The most widely used basis set for the one-particle functions of periodic
systems are plane waves. If basis sets of plane waves are employed then the
one-particle functions $\phi_{w\textbf{q}}(\textbf{r})$ are given by
\begin{equation}
\phi_{w\textbf{q}}(\textbf{r}) = \sum_\textbf{G} \,
C_{w\textbf{q}}(\textbf{G})  \; \frac{1}{\sqrt{\Omega}} \, e^{-i(\textbf{G} + \textbf{k}) \cdot \textbf{r}}. 
\label{planewaves}
\end{equation}
The elements (\ref{vx1}) of the non-local exchange potential turn into 
\begin{equation}
v_x^{NL}(\textbf{k}, \textbf{G}, \textbf{G}^{'}) =
-\frac{4\pi}{\Omega} \sum_{w\textbf{q}} \sum_{\textbf{G}^{''}}\frac{C_{w\textbf{q}}(\textbf{G}-\textbf{G}^{'}+\textbf{G}^{''})C^{*}_{w\textbf{q}}(\textbf{G}^{''})}{|\textbf{G}^{'}- \textbf{G}^{''}+ \textbf{k}-\textbf{q}|^2},
\label{Vx1insulator}
\end{equation}
and the correction term (\ref{corrHFinsulator}) turns into 
\begin{equation}
\bigg[\sum_{w} C_{w\textbf{k}}(\textbf{G})C^{*}_{w\textbf{k}}(\textbf{G}^{'})\bigg] \bigg( \tilde{F}_{\textbf{k}} - F\bigg).
\label{corrHF2insulator}
\end{equation}
Again the matrix elements $v_x^{NL}(\textbf{k}, \textbf{G}, \textbf{G}^{'})$
can be calculated by first omitting the singular
terms in Eq.\ (\ref{Vx1insulator}) and by then adding the correction term
(\ref{corrHF2insulator}).

So far we have considered only systems with bands that are either fully
occupied or fully unoccupied., i.e. we have considered isolating systems at
zero temperature. In the Appendix we sketch how the formulas of this Section
change for systems with partially filled bands and how the singularities can
be treated in this case.

\section{Results for diamond and \textit{trans}-polyacetylene}
\label{Results}

The applicability of the presented approach for treating the divergencies is now demonstrated by applying it to two cases, diamond and \textit{trans}-polyacetylene, using different approximations for the 
  exchange-correlation functionals: the Slater-Dirac (exchange-only LDA) referred to as Dirac exchange in the following, the complete exchange-correlation LDA in the parametrisation of
Vosko, Wilk, and Nusair (VWN),\cite{LDA} 
the combination of Dirac exchange plus Perdew86\cite{GGA} (P86) correlation (P86 being VWN correlation plus a gradient correction),
the EXX (exact-exchange only), and finally the combination of EXX with P86 correlation.

The pseudopotentials were generated using the pseudopotential generation code of Engel,\cite{Engel} 
  which is based on the Troullier-Martins norm-conserving scheme.\cite{Troullier} 
In all cases the pseudopotentials were generated using consistently the same functionals for exchange and correlation as for the plane wave calculations.
Relativistic effect are not included. 
The pseudopotentials of C are all constructed using a cutoff radius of 1.3 a.u.\ for both $s$ and $p$ levels.
We constructed for the calculations of \textit{trans}-polyacetylene chains hydrogen pseudopotentials with a cutoff radius of 0.9 a.u.. 
For diamond, the energy cutoff of the plane-wave basis is 60 Ry for the one-particle functions, 
and 20 Ry for the exchange potential and the response function \cite{Staedele}.
For \textit{trans}-polyacetylene,\cite{StefanEngel} we reduced the energy cutoffs to 32 Ry for the one-particle functions and 
12 Ry for the exchange potential and the response function, because we are interested
  to determine the effect of the singularity function in terms of several possible \textbf{k} points meshes, and study meshes with a large number of \textbf{k} points.

\begin{figure}[h]
   \includegraphics*[width=14.5cm]{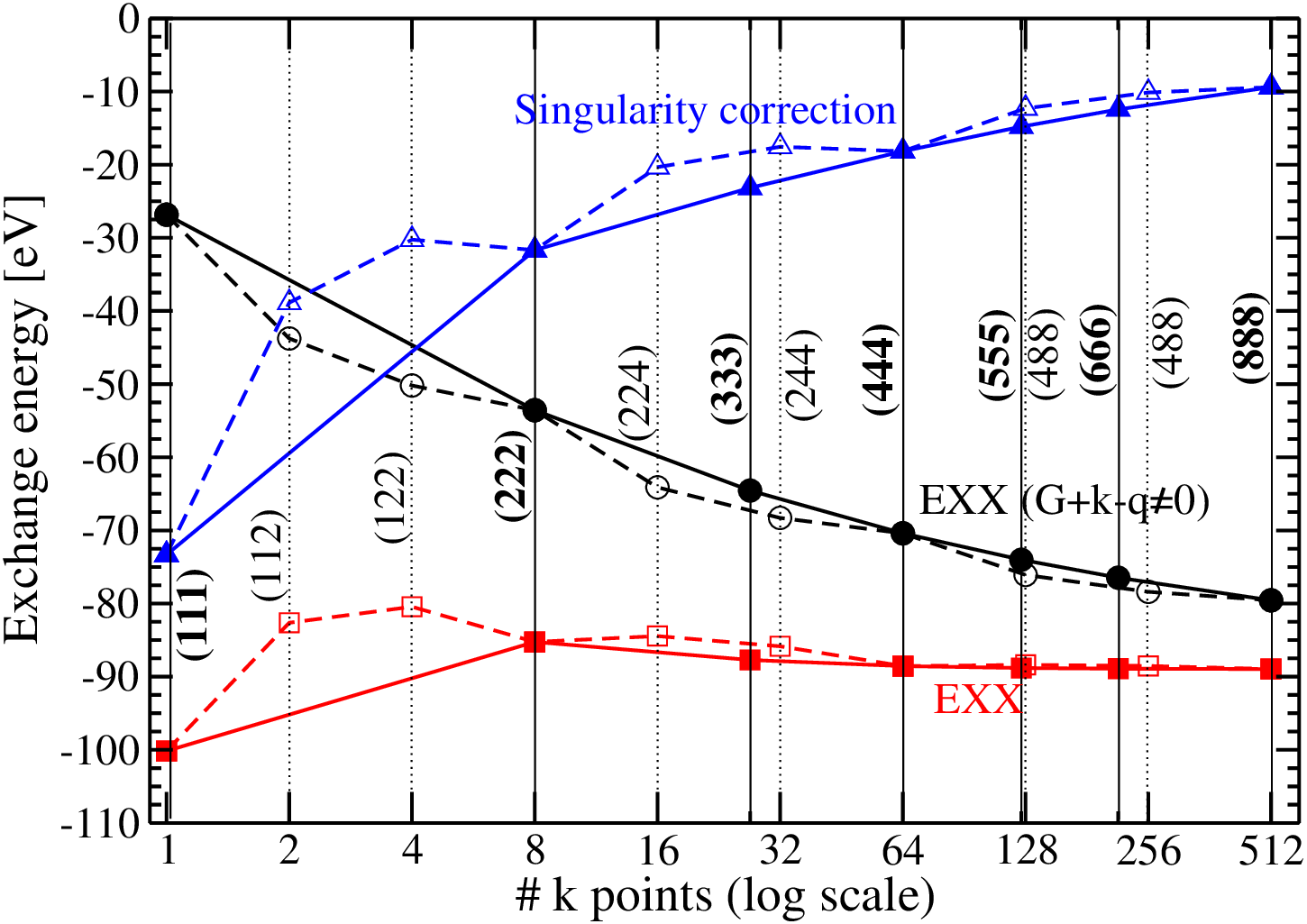}
 \caption
 { Singularity correction of the exchange energy as a function of the number of \textbf{k} points. 
   Circles represent data for the EXX energy without taking into account the singular terms.
   Triangles are the singularity correction.
   Squares constitute the full EXX energies, including the singularity correction.
   The exchange energy excluding singular terms and the singularity correction vary oppositely with the number of \textbf{k} points and their sum leads to a relatively constant and quite fast converging EXX energy.
  The lines guide the eyes for non-symmetric (dash) and symmetric (solid) \textbf{k} points meshes.
  The number of \textbf{k} points along axes of the reciprocal lattice are indicated by the numbers in parenthesis.
 }
 \label{Fig1}
 \end{figure}

\begin{figure}[h]
   \includegraphics*[width=14.5cm]{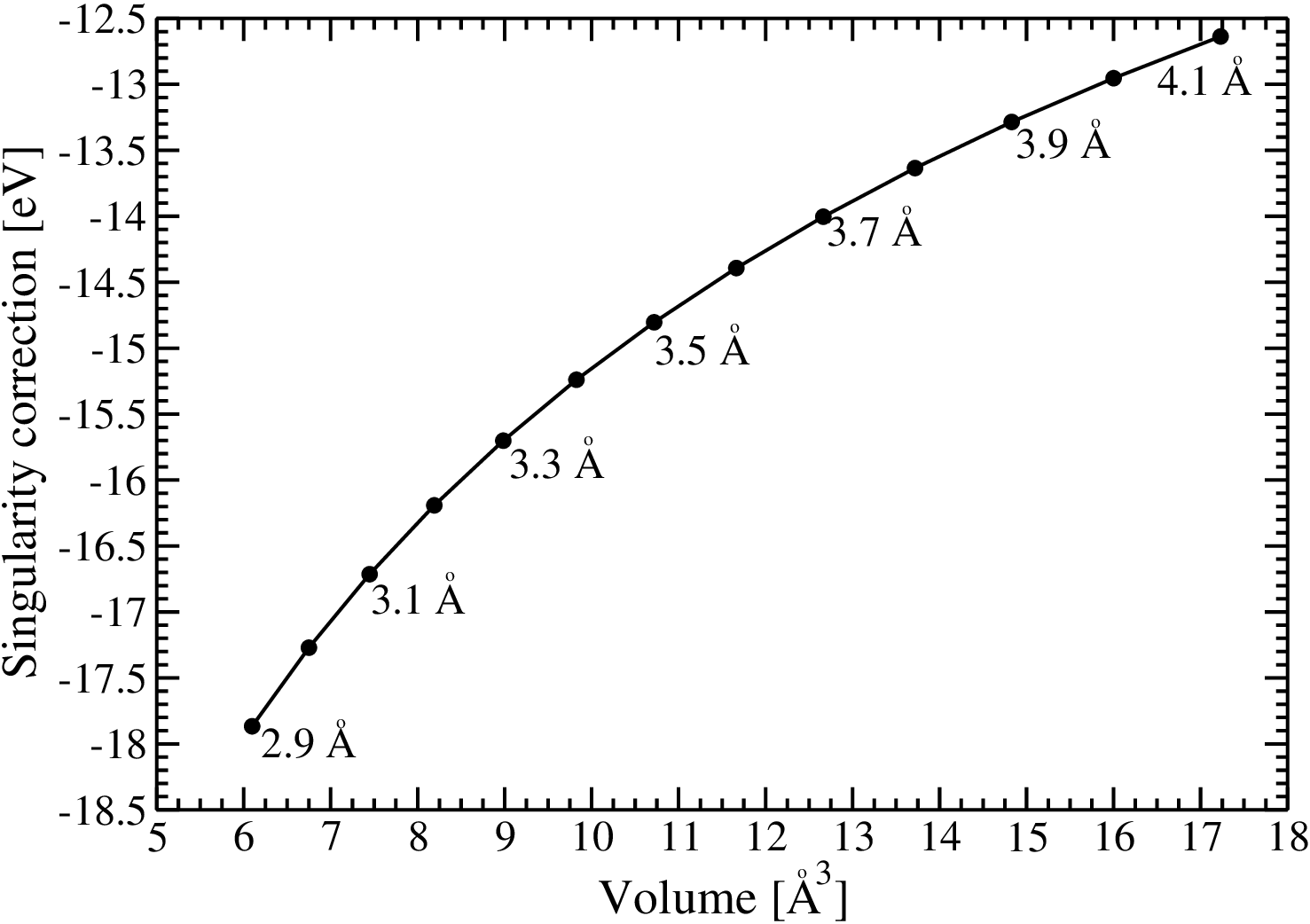}
 \caption
 { Singularity correction of the exchange energy of diamond for fixed \textbf{k} point mesh (5 $\times$ 5 $\times$ 5) as
 a function of the volume. [See text for details].
 }
 \label{Fig2}
 \end{figure}

The lattice constants for diamond were varied from 3.1 to 4.1 \AA. For comparison, the experimental lattice constant of diamond is 3.5668\AA.\cite{Madelung}
Figure \ref{Fig1} shows the variation of the singularity correction of the EXX exchange energy as a function of the \textbf{k} point mesh for diamond. 
Fig.\ \ref{Fig2} shows the singularity correction for diamond for a fixed number of \textbf{k} points but for different volumes.
Choosing 5 $\times$ 5 $\times$ 5 \textbf{k} points ensures convergence of the total energy within 0.2 eV, while 8 $\times$ 8 $\times$ 8 \textbf{k} points ensure total energy convergence within 0.05 eV. 
From Fig.\ \ref{Fig1} we notice that the singularity correction $N_v (\tilde{F} - F)$ and the exchange energy without the singularity correction vary oppositely with increasing number of \textbf{k} points.
The complete exchange energy including the singularity correction turns out to be quite stable with the number of \textbf{k} points.
Fig.\ \ref{Fig1}  also shows that for small and medium numbers of \textbf{k} points a more symmetric mesh division reduces the deviation of the complete exchange energy from its converged value at high numbers of \textbf{k} points (compare dashed with continuous EXX line).
Figure \ref{Fig2} also shows an aspect important for volume optimisations: the singularity correction changes dramatically with the volume and thus strongly 
modifies the position of the energy minimum as well as the bulk modulus.
The energy minimum is reduced because the correction function is monotonically increasing with the volume. The bulk modulus is also modified because the variation of the correction is obviously not linear. 
The bulk modulus of diamond without the singularity correction is much smaller (25\% smaller) than with it. 
Therefore, accurate integration of the singularity in the exchange energy is essential for evaluating bulk properties within EXX or HF methods.

\begin{figure}[h]
   \includegraphics*[width=14.5cm]{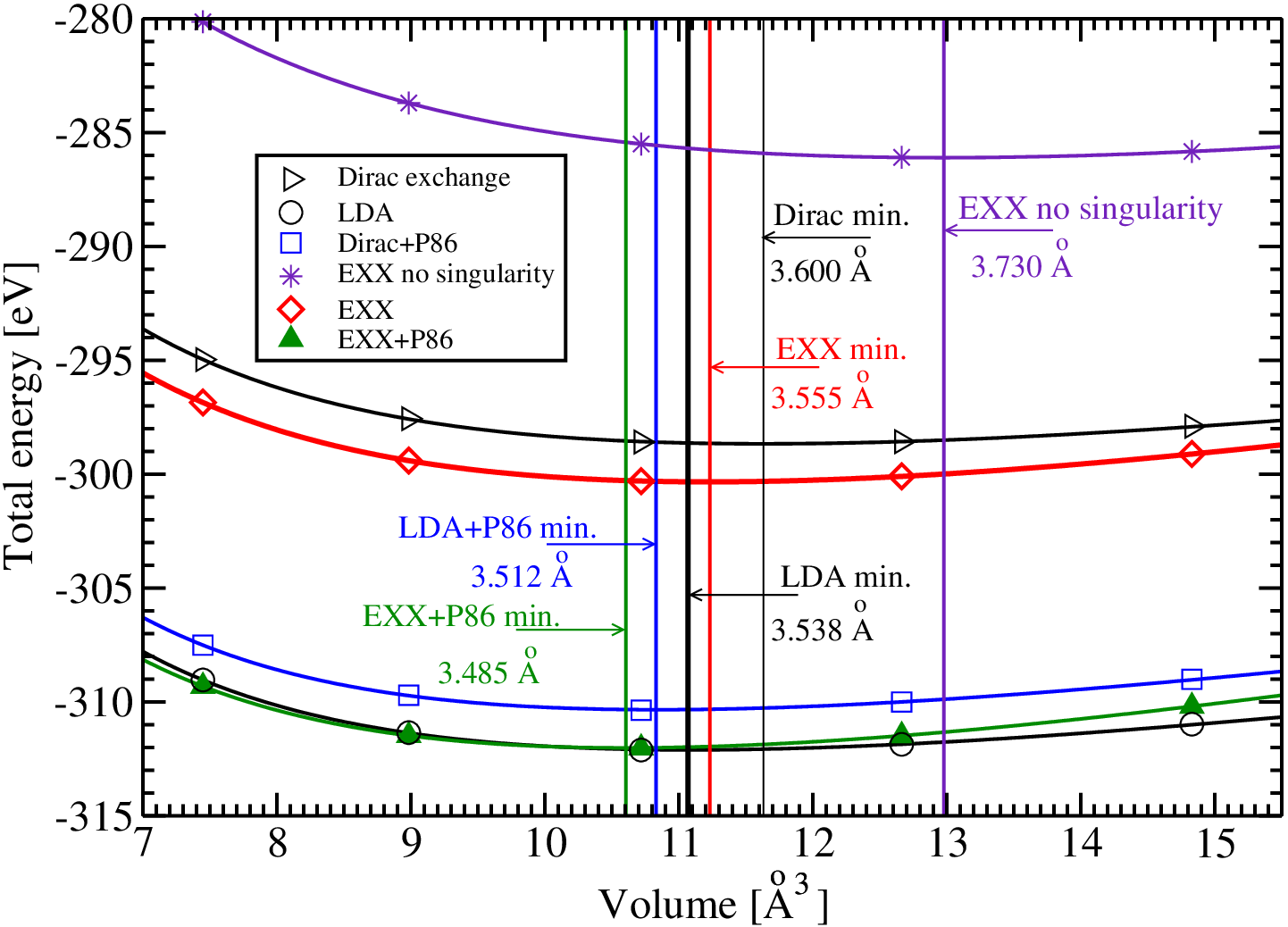}
 \caption
 { Volume optimisation for diamond. We used the equation of states of Teter.\cite{teter} 
   The total energy as a function of volume is depicted, using different exchange-correlation functionals: 
   the Dirac exchange (exchange only LDA), LDA, Dirac exchange plus P86 correlation, EXX with or without P86 correlation, and EXX without singularity correction.
   The singularity correction, as depicted in Fig.\ \ref{Fig2}, leads to a significant shift of the energy-volume minimum for EXX, 
   corresponding to a reduction of the lattice constant of -0.175 \AA.
   (Compare curves with stars and open-diamonds).
   For a discussion of correlation effects with EXX+P86 and Dirac+P86, see text.
   The experimental lattice constant of diamond is 3.5668 \AA\ (Volume = 11.3443 \AA$^3$).
 }
 \label{Fig3}
 \end{figure}

Fig.\ \ref{Fig3} shows volume optimisation results for diamond using the various combinations of exchange and correlation functionals.
We observe that a removal of Coulomb self-interactions (see EXX versus Dirac exchange) induces a significant reduction of the total energy ($\sim$2 eV). 
It also leads to a reduction of the lattice constant minimum [compare vertical lines in Fig.\ \ref{Fig3}]. 
The values of both exchange-only energy curves, i.e., of the EXX and Dirac-Slater curves, are much higher than the curves that contain a correlation potential, 
i.e., the LDA, Dirac+P86, and EXX+P86 curves, which reflects that correlation affects the total energy.
The reduction of the total energy from EXX to EXX-P86 is of the same order as the reduction from Dirac-exchange only to the Dirac exchange plus P86 correlation.  
However, the lattice constant minimum of the EXX-P86 is shifted to a much lower value than any other of the combinations of functionals.
The reason for this maybe the reintroduction of  self-interations, through the P86 correlation function. 
In any case, the poor performance of the combination EXX-P86 is not surprising because the P86 correlation is not meant 
to be used with the EXX, but rather with the LDA or GGA exchange in order to exploit error cancelations between exchange and correlation. 
Therefore, development of correlation functionals that do not depend on such error cancelations and thus are well-suited for combination 
with the EXX is highly desirable.

\begin{figure}[h]
   \includegraphics*[width=14.5cm]{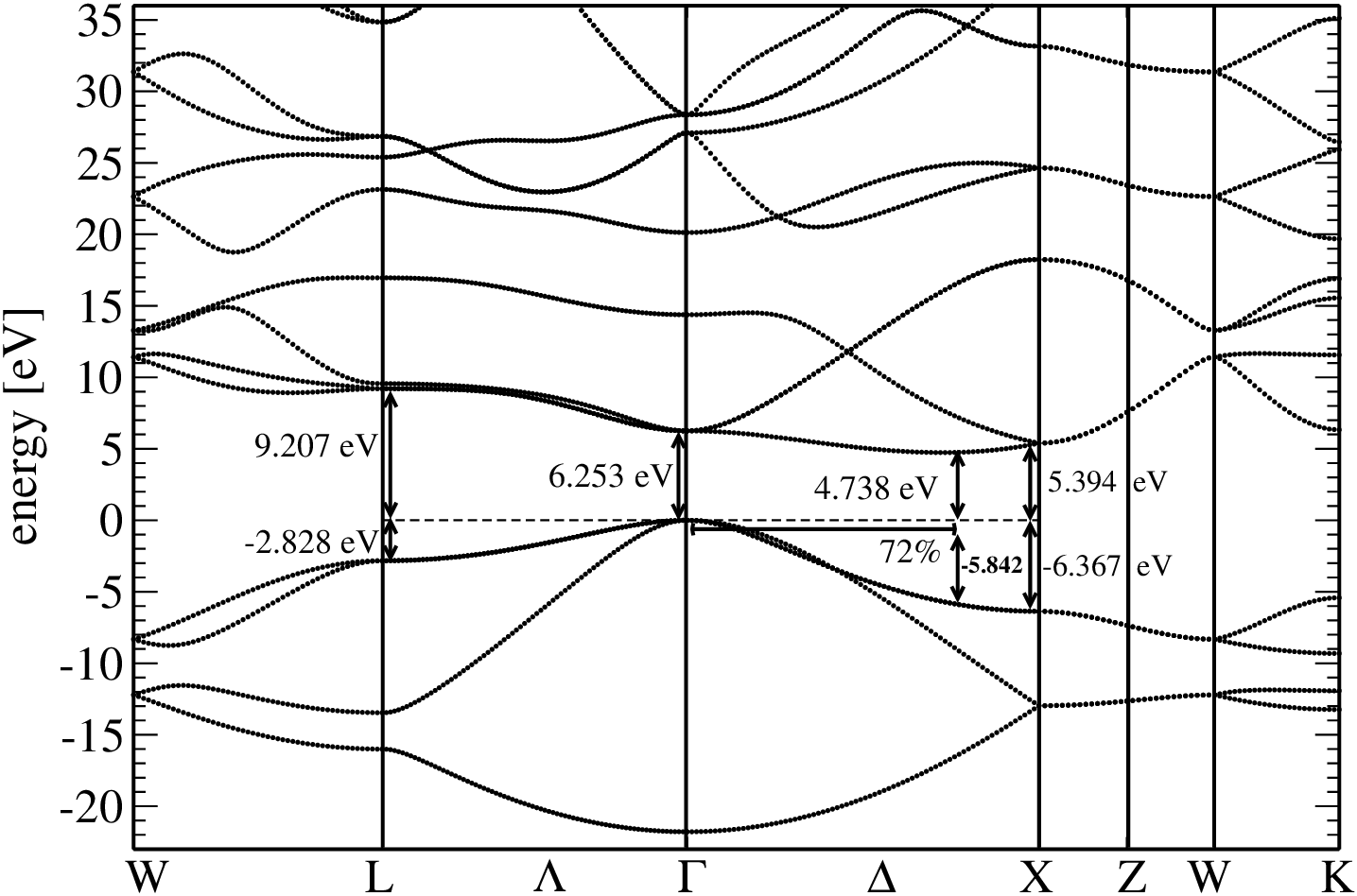}
 \caption
{
 Band structure of diamond evaluated at the EXX optimised lattice constant of 3.555 \AA [See Fig.\ \ref{Fig3}].
 The band gap of diamond is indirect, towards the $\Gamma$--$X$ direction of the BZ.
 The EXX direct transition at $\Gamma$ equals 6.253 eV. The EXX band gap equals 4.738 eV and is located at 72\% of the $X$ point away from $\Gamma$.
 The experimental lattice constant and band gap of diamond are respectively 3.5668 \AA\ and  5.50 eV.\cite{Madelung}
}
 \label{Fig4}
 \end{figure}

The EXX energy optimised lattice equals 3.555 \AA\ [see Fig.\ \ref{Fig3}]. 
It becomes natural now to evaluate the band structure at the EXX energy minimum.
Figure  \ref{Fig4} shows the band structure at this EXX energy minimum.
The indirect EXX bandgap is 4.838 \AA, 
a value comparable to previous published data,\cite{Staedele}
and much closer to the experimental band gap of 5.50 eV\cite{Madelung} than the LDA value of 3.90 eV.\cite{Hybertsen} 
The experimental lattice constant of diamond\cite{Madelung} (3.5668 \AA) is slightly larger (+0.011 \AA) than the EXX energy optimised lattice.
Going from the EXX lattice minimum to the experimental value, i.e., addition of 0.011 \AA\ to the EXX lattice constant, 
leads to minute reduction of the band gap.
However, we want to emphasize that the singularity correction to the exchange energy is obviously essential for determining the right correspondence between the EXX energy lattice minimum [Fig.\ \ref{Fig3}]
and the band structure or its band gap [Fig.\ \ref{Fig4}]. 
For instance, without the singularity correction the EXX lattice energy minimum would be incorrectly 
overestimated (3.730 \AA\ instead of 3.555 \AA\ as illustrated in Fig.\ \ref{Fig3}), 
and the EXX band gap correspondingly would be largely underestimated (because the band gap variation is roughly inversely proportional to the lattice constant).
In summary, we find that for diamond the EXX band gap is somewhat smaller than in the experiments (0.66 eV lower than experiment), but EXX improves significantly the LDA value ($\sim$1.6 eV lower than experiment).
As mentioned in the introduction, the EXX calculation does not account for the derivative discontinuity\cite{PerdewDISC, Sham} of the band gap at 
integer electron numbers and, of course, also not for the correlation potential.

\begin{figure}[h]
   \includegraphics*[width=14.5cm]{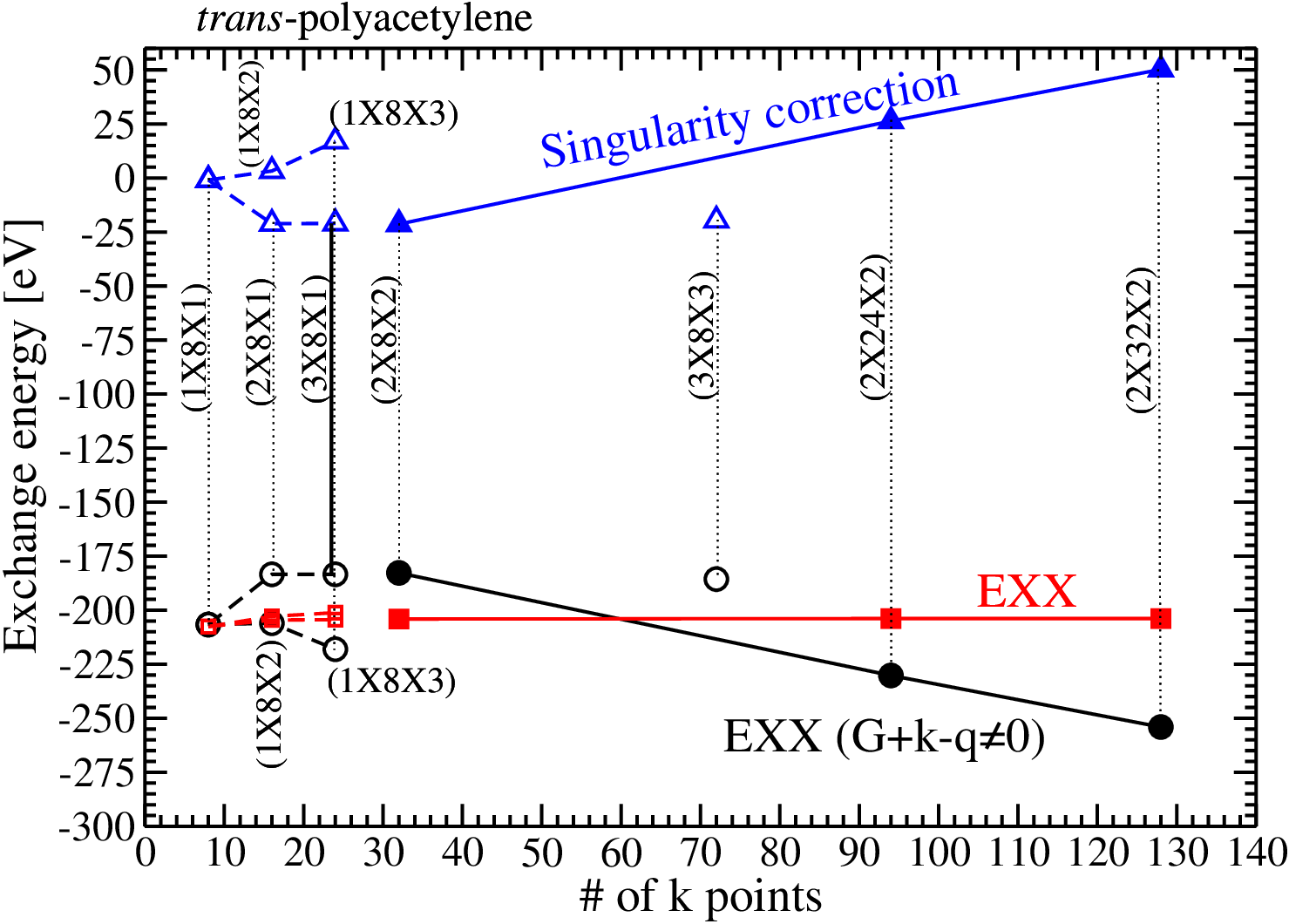}
 \caption
 { Singularity correction of EXX exchange energy of \textit{trans}-polyacetylene as a function of the number of \textbf{k} points. 
  See caption of Fig.\ \ref{Fig1} for symbols description.
  The number of \textbf{k} points along axes of the reciprocal lattice are indicated by the numbers in parenthesis.
  The second entry refers to the number of \textbf{k} points along the \textit{trans}-polyacetylene chain.
 }
 \label{Fig5}
 \end{figure}

We now consider a more general crystal structure: \textit{trans}-polyacetylene.
The unit cell is constituted of 4 carbons and 4 hydrogens.
More data on the band structure of \textit{trans}-polyacetylene can be found elsewhere.\cite{StefanEngel}
\textit{Trans}-polyacetylene constitutes a monoclinic lattice structure (group $P2_1/a$) with the following lattice parameters expressed in cartesian
coordinates (and in \AA):
\begin{eqnarray}
\textbf{a}_1 & = & (4.24, 0.00, 0.00) \nonumber \\
\textbf{a}_2 & = & (-0.0642644, 2.454158, 0.00) \\
\textbf{a}_3 & = & (0.00, 0.00, 7.32)\nonumber
\end{eqnarray}
The angle between $\textbf{a}_1$ and $\textbf{a}_2$ is 91.46$^{\circ}$. 
The angle between any two dimerised chains is 55$^{\circ}$.
The coordinates of the structurally optimised hydrogen atoms come from Hartree-Fock calculations
and the lattice parameters and C--C bond distances and angles come from experimental values.\cite{Hartree}
This structure constitutes a general and realistic case to test our singularity correction for several \textbf{k} points meshes. 
A graph equivalent to Fig.\ \ref{Fig1} is displayed in Fig.\ \ref{Fig5} for this molecular crystal.

Figure \ref{Fig5} shows for \textit{trans}-polyacetylene a similar trend as shown in Fig.\ \ref{Fig1} for diamond. 
That is, the singularity correction and the exchange energy excluding the singularity vary in opposite matter, for any chosen \textbf{k} points division. 
The complete exchange energy is a rather monotonic function of the \emph{total} number of \textbf{k} points in the unit cell. 

The results, both for diamond and \textit{trans}-polyacetylene, show that the approach presented here to treat the integrable singularities in the KS and HF methods constitutes a stable and general scheme.

\section{Concluding remarks}

\label{Conclude}

We have presented a general scheme for treating the integrable singularities of the exchange energy within the EXX or the HF formalisms.
We have shown that the divergent terms in the exchange energy depend only on the number and positions of \textbf{k} points and on the unit cell vectors and thus on the unit cell volume,
but not on the single particle wave functions or on the particular atomic positions within the unit cell.
A similar correction procedure is proposed for matrix elements of the non-local exchange operator which occurs in the Hartree-Fock methods and can be used to construct the exact local Kohn-Sham exchange potential.
We applied the singularity correction to a typical symmetric structure, diamond, and to a more general structure, \textit{trans}-polyacetylene, and discussed
the effect of the singularity function on volume optimisation and \textbf{k} points convergence.
The singularity function depends strongly on the total number of \textbf{k} points and more weakly on the choice of the specific 
division of the \textbf{k} points mesh.
The complete, i.e., singularity corrected exchange energy, converges well with the number of \textbf{k} points.
The method proposed here constitutes a stable, simple to implement, and general scheme that can be applied to systems with 
any lattice parameters within either the EXX Kohn-Sham or the Hartree-Fock formalism.


\section{Acknowledgments}

This work was supported by the \textit{Alexander von-Humboldt Stiftung} (P.\ Carrier) and by the \textit{Deutsche Forschungsgemeinschaft} (DFG).

\newpage
\section{Appendix}

In this Appendix we briefly consider the treatment of singularities for systems
with partially filled bands. In this case expression (\ref{ExSigma}) for the exchange
energy turns into
\begin{eqnarray}
E_x = 
-\frac{1}{N_k} \sum_{v\textbf{k}}^{\mbox{occ.}}
\sum_{w\textbf{q}}^{\mbox{occ.}} \eta_{v\textbf{k}} \, \eta_{w\textbf{q}} \int_{\Omega} \! d\textbf{r} 
\int_{\Omega} \! d\textbf{r}^{'} \,
\frac{\phi^{\dag}_{v\textbf{k}}(\textbf{r}) \phi_{w\textbf{q}}(\textbf{r})
      \phi^{\dag}_{w\textbf{q}}(\textbf{r}^{'}) \phi_{v\textbf{k}}(\textbf{r}^{'})}
{|\textbf{r}-\textbf{r}^{'}|}, 
\label{ExSigma2}
\end{eqnarray}
with the occupation factor $\eta_{v\textbf{k}}$. The occupation factor shall not
include the spin multiplicity, i.e., $0 \le \eta_{v\textbf{k}} \le 1$. 
Analogously to Eq.\ (\ref{ex1}) the exchange energy can be expressed by
\begin{equation}
E_x  =
-\frac{4\pi}{N_k\Omega} \sum_{v\textbf{k}} \sum_{w\textbf{q}} \eta_{v\textbf{k}} \, \eta_{w\textbf{q}} \sum_{\textbf{G}}
\frac{Y^{*}_{w\textbf{q},v\textbf{k}}(\textbf{G}) \, 
Y_{w\textbf{q},v\textbf{k}}(\textbf{G})}{|\textbf{G}+\textbf{k}-\textbf{q}|^2},
\label{ex1partial}
\end{equation}
The singular terms in expression (\ref{ex1partial}), namely those with $\textbf{G}=\textbf{0}$, $\textbf{k} = \textbf{q}$, and $v=w$, 
can be treated in analogy to Eq.\ (\ref{singularterms2}) according to 
\begin{eqnarray}
&& \!\!\!\!\!\!\!\!-\frac{4\pi}{N_k\Omega} \frac{\Omega^2}{(2\pi)^6} \sum_{v}  \int_{BZ} \! d\textbf{k} \,
\int_{BZ} \! d\textbf{q} \; \eta_{v\textbf{q}} \, \eta_{v\textbf{k}} \,
\frac{Y^{*}_{v\textbf{q},v\textbf{k}}(\textbf{0}) \, 
Y_{v\textbf{q},v\textbf{k}}(\textbf{0})}{|\textbf{k}-\textbf{q}|^2}
\nonumber \\
&=& -\frac{4\pi}{N_k\Omega} \frac{\Omega^2}{(2\pi)^6} \sum_{v} \int_{BZ} \! d\textbf{k} \,
 \int_{BZ} \! d\textbf{q} \, \bigg[ \eta_{v\textbf{q}} \, \eta_{v\textbf{k}} \,
\frac{Y^{*}_{v\textbf{q},v\textbf{k}}(\textbf{0}) \, 
Y_{v\textbf{q},v\textbf{k}}(\textbf{0})}{|\textbf{k}-\textbf{q}|^2} 
-\eta_{v\textbf{k}}^2 \, f(\textbf{k}-\textbf{q}) \bigg] 
\nonumber \\
&& -\frac{4\pi}{N_k\Omega} \frac{\Omega^2}{(2\pi)^6} \sum_{v} \int_{BZ} \! d\textbf{k} \,
 \int_{BZ} \! d\textbf{q} \; \eta_{v\textbf{k}}^2 \, f(\textbf{k}-\textbf{q})
\nonumber \\
&\approx& -\frac{4\pi}{N_k\Omega} \sum_{v} \sum_{\textbf{k}}
\sum_{\textbf{q} \ne \textbf{k}} \bigg[ \eta_{v\textbf{q}} \, \eta_{v\textbf{k}} \,
\frac{Y^{*}_{v\textbf{q},v\textbf{k}}(\textbf{0}) \, 
Y_{v\textbf{q},v\textbf{k}}(\textbf{0})}{|\textbf{k}-\textbf{q}|^2} -
\eta_{v\textbf{k}}^2 \, f(\textbf{k}-\textbf{q}) \bigg]
\nonumber \\
&& -\frac{4\pi}{N_k\Omega} \frac{\Omega^2}{(2\pi)^6} \sum_{v} \int_{BZ} \!
d\textbf{k} \, \eta_{v\textbf{k}}^2 \,
 \int_{BZ} \! d\textbf{q} \; f(\textbf{q})
\nonumber \\
&=& -\frac{4\pi}{N_k\Omega} \sum_{v} \sum_{\textbf{k}}
\sum_{\textbf{q} \ne \textbf{k}} \eta_{v\textbf{q}} \, \eta_{v\textbf{k}} \,
\frac{Y^{*}_{v\textbf{q},v\textbf{k}}(\textbf{0}) \, 
Y_{v\textbf{q},v\textbf{k}}(\textbf{0})}{|\textbf{k}-\textbf{q}|^2}
\nonumber \\
&& + \frac{1}{N_k} \sum_{\textbf{k}}  \sum_{v} \eta_{v\textbf{k}}^2 \,
\frac{4\pi }{\Omega} \sum_{\textbf{q} \ne \textbf{k}} f(\textbf{k}-\textbf{q}) 
\;-\; \left [ \frac{\Omega}{(2\pi)^3 N_k} \sum_{v} \int_{BZ} \!
d\textbf{k} \, \eta_{v\textbf{k}}^2 \right ] \left [\frac{4\pi}{(2\pi)^3} \,
 \int_{BZ} \! d\textbf{q} \; f(\textbf{q}) \right ]
\nonumber \\
&\approx& -\frac{4\pi}{N_k\Omega} \sum_{v} \sum_{\textbf{k}}
\sum_{\textbf{q} \ne \textbf{k}} \eta_{v\textbf{q}} \, \eta_{v\textbf{k}} \,
\frac{Y^{*}_{v\textbf{q},v\textbf{k}}(\textbf{0}) \, 
Y_{v\textbf{q},v\textbf{k}}(\textbf{0})}{|\textbf{k}-\textbf{q}|^2}
\nonumber \\
&& + \frac{1}{N_k}  \sum_{v} \sum_{\textbf{k}} \eta_{v\textbf{k}}^2 \,
 \tilde{F}_{\textbf{k}}
\;-\; \left [ \frac{1}{N_k} \sum_{v} \sum_{\textbf{k}} \, \eta_{v\textbf{k}}^2 \right ] \left [\frac{4\pi}{(2\pi)^3} \,
 \int_{BZ} \! d\textbf{q} \; f(\textbf{q}) \right ]
\nonumber \\
&=& -\frac{4\pi}{N_k\Omega} \sum_{v} \sum_{\textbf{k}}
\sum_{\textbf{q} \ne \textbf{k}} \eta_{v\textbf{q}} \, \eta_{v\textbf{k}} \,
\frac{Y^{*}_{v\textbf{q},v\textbf{k}}(\textbf{0}) \, 
Y_{v\textbf{q},v\textbf{k}}(\textbf{0})}{|\textbf{k}-\textbf{q}|^2}
\nonumber \\
&& + \frac{1}{N_k} \sum_{v} \sum_{\textbf{k}}  \eta_{v\textbf{k}}^2 \,
 \tilde{F}_{\textbf{k}}
\;-\; \left [ \frac{1}{N_k} \sum_{v} \sum_{\textbf{k}} \, \eta_{v\textbf{k}}^2 \right ] F,
\label{singularterms2partial}
\end{eqnarray}
with $\tilde{F}_{\textbf{k}} = \frac{4\pi}{\Omega} \sum_{\textbf{q}\ne \textbf{k}} f({\textbf{q} - \textbf{k}})$ 
and $F = - \frac{4\pi}{(2\pi)^3} \int_{BZ}\! d\textbf{q}\;  f(\textbf{k} - \textbf{q})$. In
Eq. (\ref{singularterms2partial}) we have used that the integral 
$\int_{BZ} \! d\textbf{k} \, \eta_{v\textbf{k}}^2$ does not contain any
singularities and therefore can be evaluated via
summation over the $\textbf{k}$ points. The energy again can be evaluated by
first omitting the singular terms in Eq.\ (\ref{ex1partial}) and by then
adding the correction term
\begin{eqnarray}
\frac{1}{N_k} \sum_{\textbf{k}}  \sum_{v} \eta_{v\textbf{k}}^2 \,
 \tilde{F}_{\textbf{k}}
\;-\; \left [ \frac{1}{N_k} \sum_{v} \sum_{\textbf{k}} \, \eta_{v\textbf{k}}^2 \right ] F.
\label{correctionpartial}
\end{eqnarray}
For the particular case of a uniform grid of $\textbf{k}$ points the functions
$\tilde{F}_{\textbf{k}}$ all equal the function $\tilde{F}$ of Eq.\
(\ref{Finteg}) and the correction term turns into  
\begin{eqnarray}
\left [ \frac{1}{N_k} \sum_{v} \sum_{\textbf{k}} \, \eta_{v\textbf{k}}^2
\right ] [\tilde{F} - F] \,.
\label{correction partial1}
\end{eqnarray}

In a similar way as the treatment of the singularities in the exchange energy was generalized for the case of partially occupied bands also the treatment
of the singularities in the exchange potential can be generalized to the case of partially occupied bands. The matrix elements of the non-local 
exchange potential, $v_x^{NL}(\textbf{k},\mu, \nu)$, then are given by
\begin{equation}
v_x^{NL}(\textbf{k}, \mu, \nu) =
-\frac{4\pi}{\Omega}  \sum_{w\textbf{q}} \eta_{w\textbf{q}} \sum_{\textbf{G}}
\frac{Y^{*}_{w\textbf{q},\mu\textbf{k}}(\textbf{G}) \, 
Y_{w\textbf{q},\nu\textbf{k}}(\textbf{G})}{|\textbf{G}+\textbf{k}-\textbf{q}|^2},
\label{vx1partial}
\end{equation}
Expression (\ref{vx1partial}) contains singular terms, i.e., those with
$\textbf{G}=\textbf{0}$ and $\textbf{k} = \textbf{q}$. In the limit of an
infinite number of unit cells the summation over $\textbf{q}$ again turns into
an integral, namely
\begin{eqnarray}
 v_x^{NL}(\textbf{k}, \mu, \nu)=
-\frac{4\pi}{\Omega} \frac{\Omega}{(2\pi)^3} \, 
\sum_{w} \int_{BZ} \! d\textbf{q} \, \eta_{w\textbf{q}} \,\sum_{\textbf{G}}
\frac{Y^{*}_{w\textbf{q},\mu\textbf{k}}(\textbf{G}) \, 
Y_{w\textbf{q},\nu\textbf{k}}(\textbf{G})}{|\textbf{G}+\textbf{k}-\textbf{q}|^2}.
\label{vx2}
\end{eqnarray}
with an integrable singularity. We can now treat the singular terms in the
right hand side of (\ref{vx2}) exactly analogously to the singular terms
occuring in the exchange energy:
\begin{eqnarray}
&& \!\!\!\!\!\!\!\!-\frac{4\pi}{\Omega}  \frac{\Omega}{(2\pi)^3} \sum_{w}
\int_{BZ} \! d\textbf{q} \,\, \eta_{w\textbf{q}} \,
\frac{Y^{*}_{w\textbf{q},\mu\textbf{k}}(\textbf{0}) \, 
Y_{w\textbf{q},\nu\textbf{k}}(\textbf{0})}{|\textbf{k}-\textbf{q}|^2}
\nonumber \\
&=& -\frac{4\pi}{\Omega} \frac{\Omega}{(2\pi)^3} 
\sum_{w}  \int_{BZ} \! d\textbf{q} \, \bigg[ \eta_{w\textbf{q}} \,
\frac{Y^{*}_{w\textbf{q},\mu\textbf{k}}(\textbf{0}) \, 
Y_{w\textbf{q},\nu\textbf{k}}(\textbf{0})}{|\textbf{k}-\textbf{q}|^2} 
\, - \, \eta_{w\textbf{k}} \, Y^{*}_{w\textbf{k},\mu\textbf{k}}(\textbf{0}) \, 
Y_{w\textbf{k},\nu\textbf{k}}(\textbf{0})  \, f(\textbf{k}-\textbf{q}) \bigg] 
\nonumber \\
&& - \, \left [ \sum_{w}  \eta_{w\textbf{k}} \, Y^{*}_{w\textbf{k},\mu\textbf{k}}(\textbf{0}) \, 
Y_{w\textbf{k},\nu\textbf{k}}(\textbf{0}) \right ] \,\frac{4\pi}{\Omega} \frac{\Omega}{(2\pi)^3} 
 \int_{BZ} \! d\textbf{q} \; f(\textbf{k}-\textbf{q})
\nonumber \\
&\approx& -\frac{4\pi}{\Omega} \sum_{w}
\sum_{\textbf{q} \ne \textbf{k}} \bigg[ \eta_{w\textbf{q}} \,
\frac{Y^{*}_{w\textbf{q},\mu\textbf{k}}(\textbf{0}) \, 
Y_{w\textbf{q},\nu\textbf{k}}(\textbf{0})}{|\textbf{k}-\textbf{q}|^2} - \,
\eta_{w\textbf{k}} \, Y^{*}_{w\textbf{k},\mu\textbf{k}}(\textbf{0}) \, 
Y_{w\textbf{k},\nu\textbf{k}}(\textbf{0}) \,
f(\textbf{k}-\textbf{q}) \bigg]
\nonumber \\
&& - \, \left [ \sum_{w} \eta_{w\textbf{k}} \, Y^{*}_{w\textbf{k},\mu\textbf{k}}(\textbf{0}) \, 
Y_{w\textbf{k},\nu\textbf{k}}(\textbf{0}) \right ] \,\frac{4\pi}{\Omega} \frac{\Omega}{(2\pi)^3} 
 \int_{BZ} \! d\textbf{q} \; f(\textbf{q})
\nonumber \\
&=& -\frac{4\pi}{\Omega} \sum_{w} 
\sum_{\textbf{q} \ne \textbf{k}}  \eta_{w\textbf{q}} \,
\frac{Y^{*}_{w\textbf{q},\mu\textbf{k}}(\textbf{0}) \, 
Y_{w\textbf{q},\nu\textbf{k}}(\textbf{0})}{|\textbf{k}-\textbf{q}|^2}
\, + \, \left [ \sum_{w} \eta_{w\textbf{k}} \, Y^{*}_{w\textbf{k},\mu\textbf{k}}(\textbf{0}) \, 
Y_{w\textbf{k},\nu\textbf{k}}(\textbf{0}) \right ] \, [\tilde{F}_\textbf{k} - F].
\label{singularterms3partial}
\end{eqnarray}

Thus, the matrix elements $v_x^{Nl}(\textbf{k},\mu,\nu)$ of the non-local exchange potential
can be evaluated by first omitting the singular
terms in 
Eq.\ (\ref{vx1partial}) and by then adding the
following correction term
\begin{equation}
\bigg[\sum_{w} \eta_{w\textbf{k}} \, Y^{*}_{w\textbf{k},\mu\textbf{k}}(\textbf{0}) \, 
Y_{w\textbf{k},\nu\textbf{k}}(\textbf{0}) \bigg] \bigg( \tilde{F}_{\textbf{k}} - F\bigg).
\label{corrHF}
\end{equation}
The required sums $\tilde{F}_{\textbf{k}}$ and the integral $F$ have to be calculated only \emph{once} at the beginning of the self-consistency procedure and then multiplied by 
$\sum_{w} \eta_{w\textbf{k}} \, Y^{*}_{w\textbf{k},\mu\textbf{k}}(\textbf{0}) \, 
Y_{w\textbf{k},\nu\textbf{k}}(\textbf{0})$ in each HF self-consistency cycle, because the $Y^{*}_{w\textbf{k},\mu\textbf{k}}(\textbf{0}) \, 
Y_{w\textbf{k},\nu\textbf{k}}(\textbf{0})$ and the occupation numbers $\eta_{w\textbf{k}}$ change
\emph{during} the self-consistency cycle. 
In case of a uniform grid of $\textbf{k}$ points the $\tilde{F}_{\textbf{k}}$ reduce to $\tilde{F}$ given in Eq.\ (\ref{uniform}), 
as for fully occupied bands (since the singularity function $f$(\textbf{q}) does not depend on the occupation numbers $\eta_{w\textbf{k}}$).

If basis sets of plane waves are employed then the elements (\ref{vx1partial}) of the non-local exchange potential turn into 
\begin{equation}
v_x^{NL}(\textbf{k}, \textbf{G}, \textbf{G}^{'}) =
-\frac{4\pi}{\Omega} \sum_{w\textbf{q}}  \eta_{w\textbf{q}} \, \sum_{\textbf{G}^{''}}\frac{C_{w\textbf{q}}(\textbf{G}-\textbf{G}^{'}+\textbf{G}^{''})C^{*}_{w\textbf{q}}(\textbf{G}^{''})}{|\textbf{G}^{'}- \textbf{G}^{''}+ \textbf{k}-\textbf{q}|^2},
\label{Vx1}
\end{equation}
and the correction term (\ref{corrHF}) turns into 
\begin{equation}
\bigg[\sum_{w} \eta_{w\textbf{k}} \, C_{w\textbf{k}}(\textbf{G})C^{*}_{w\textbf{k}}(\textbf{G}^{'})\bigg] \bigg( \tilde{F}_{\textbf{k}} - F\bigg).
\label{corrHF2}
\end{equation}
Again the matrix elements $v_x^{NL}(\textbf{k}, \textbf{G}, \textbf{G}^{'})$
can be calculated by first omitting the singular
terms in Eq.\ (\ref{Vx1}) and by then adding the correction term
(\ref{corrHF2}).

\newpage

\newpage

 \newpage

 \newpage

 \newpage


\begin{thebibliography}{32}
\bibitem{note60}
{With $N=60$ the singularity correction typically is converged to below 5 meV; For the results given in this paper $N=120$ was used, corresponding to an accuracy below 1 meV.}
\end{thebibliography}
\end{document}